\documentclass[aps,twocolumn ]{revtex4}
\usepackage{amsfonts}
\usepackage{amsmath}
\usepackage{amssymb,epsf}

\begin{document}

\title{Alternative approach to thermodynamic phase transitions}
\author{Seyed Hossein Hendi$^{1,2}$\footnote{
email address: hendi@shirazu.ac.ir}, Shahram Panahiyan$^{1,3,4}$
\footnote{
email address: shahram.panahiyan@uni-jena.de}, Behzad Eslam Panah$^{1,2,5}$%
\footnote{
email address: behzad.eslampanah@gmail.com} and Mubasher Jamil$^{6,7}$\footnote{%
email address: mjamil@zjut.edu.cn}}
\affiliation{\vspace{0.3cm}$^1$ Physics Department and Biruni Observatory, College of
Sciences, Shiraz University, Shiraz 71454, Iran\\
$^2$ Research Institute for Astronomy and Astrophysics of Maragha (RIAAM),
P. O. Box 55134-441, Maragha, Iran\\
$^3$ Physics Department, Shahid Beheshti University, Tehran 19839, Iran\\
$^4$ Helmholtz-Institut Jena, Fr\"{o}belstieg 3, Jena 07743, Germany\\
$^5$ ICRANet, Piazza della Repubblica 10, I-65122 Pescara, Italy\\
$^6$ School of Natural Sciences (SNS), National University of Sciences and
Technology, H-12, Islamabad, 44000, Pakistan\\
$^7$ Institute for Astrophysics, Zhejiang University of Technology, Hangzhou, China}

\begin{abstract}
One of the major open problems in theoretical physics is a
consistent quantum gravity theory. Recent developments in
thermodynamic phase transitions of black holes and their van der
Waals-like behavior may provide an interesting quantum
interpretation of classical gravity. Studying different methods of
investigating phase transitions can extend our insight into the
nature of quantum gravity. In this paper, we present an
alternative theoretical approach for finding thermodynamic phase
transitions in the extended phase space. Unlike the standard
methods based on the usual equation of state involving
temperature, our approach uses a new quasi-equation constructed
from the slope of temperature versus entropy. This approach
addresses some of the shortcomings of the other methods, and
provides a simple and powerful way of studying the critical
behavior of a thermodynamical system. Among the applications of
this approach, we emphasize the analytical demonstration of
possible phase transition points, and the identification of the
non-physical range of horizon radii for black holes.
\end{abstract}

\maketitle

\section{Introduction\label{INT}}

Phase transitions are one of the interesting and challenging
experimental and theoretical topics in various areas of science,
from biological \cite{bio} to physical systems \cite{nice}.
Regarding the physical systems, one may find that phase
transitions play an important role in elementary particle physics
\cite{El-Par}, condensed matter
\cite{Con-MatI,Con-MatII,Con-MatIII}, the usual thermodynamics
\cite{Thermo}, cosmology \cite{Cosmol}, black holes \cite{BHs} and
other branches of physics. In general, there are at least three
well-known approaches to discuss phase transitions inside a black
hole. Two of these approaches are based on a macroscopic point of
view (Davies and Landau-Lifshitz methods that discuss,
respectively, the behavior of heat capacity and thermodynamic
fluctuations), and one of them is based on a microscopic viewpoint
(thermodynamic geometry or Ruppeiner geometry). In this regard, we
focus on the thermodynamic phase transition of black holes. There
are several well-known approaches for studying the critical
behavior of black holes. One of them uses heat capacity in the
context of canonical ensemble. The discontinuity of heat capacity
is where the phase transition takes place. Another method is based
on studying the van der Waals-like behavior of black holes in the
extended phase space by considering a proportionality between the
cosmological constant and dynamical pressure. This method is based
on an equation of state, which originates from temperature. For
details, please refer to earlier reviews on phase transitions of
black holes, black rings, black saturns and black membranes
\cite{r1,r2,r3, r4}, where classical instability and horizon
topology changing transitions are also discussed.

In this paper, we consider the method for extracting the critical
values and the van der Waals phase transition based on the
divergence of heat capacity (the Davies method). We should note
that our method is applicable in the extended phase space
thermodynamics of black hole, in which the cosmological constant
is considered as dynamic pressure. Due to the fact that in our
approach we use the behavior of the heat capacity, it is not a
novel approach, but an alternative way for obtaining the
properties of heat capacity based on the canonical ensemble. In
other words, this method is the same as the Davies method, but
with a different point of view. We introduce an alternative
approach for obtaining the critical values of the van der
Waals-like behavior by using the slope of $T$ versus $S$, and give
two relevant examples. In other words, in the usual extended phase
space, one calculates $T=\left( \frac{\partial M}{\partial S}
\right) $ to obtain an equation of state. Other calculations and
interpretations are based on such an equation of state. In our
approach, we use $\left( \frac{\partial T}{\partial S}\right)
=\left( \frac{\partial ^{2}M}{\partial S^{2}}\right)=0 $ to obtain
a new relation for pressure. This relation is free of temperature,
and therefore, it is not the usual equation of state. One finds
that this relation, and other related quantities and phase
diagrams, are different from those obtained in the usual extended
phase space. Our method provides the possibility of mapping all
phase transition points in a system. In addition, it can single
out the non-physical range of horizon radii in which the black
hole solutions do not exist. Furthermore, the rate of increase of
non-physical range of horizon radii for different critical values
can be obtained. As a final point, we should note that our
approach is quite different from the Poincare's turning point
method \cite{PoincareI,PoincareII,PoincareIII,PoincareIV}, which
is a powerful tool for investigating the turning point and
stability. Unlike this method, our approach is based on the
extended phase space thermodynamics and on the slope of $T$ versus
$S$ as a primitive equation for obtaining the dynamical pressure,
where its maximum is related to a critical point and possible
phase transition.

The layout of the paper is as follows: In Sec. \ref{Old}, we give
a brief review of the canonical ensemble approach and the van der
Waals approach to black hole phase transitions. In Sec.
\ref{NewProp}, we introduce an alternative scheme for calculating
the phase transition parameters, and we employ this method to the
usual van der Waals system. In Sec. \ref{RN}, we perform a case
study of the Reissner-Nordstr\"{o}m AdS black hole to elaborate
the method. Finally, we make several remarks in the
conclusion.\newline

\section{Brief review of various methods \label{Old}}

\subsection{The usual method: heat capacity and extended phase space}

In the canonical ensemble, discontinuities of the heat capacity
are indicated as the phase transition points. The heat capacity in
the context of canonical ensemble is given by
\begin{equation}
C_{Q}=\frac{T}{\left( \frac{\partial ^{2}M}{\partial S^{2}}\right) _{Q}}%
=T\left( \frac{\partial S}{\partial T}\right) _{Q}.  \label{CQ}
\end{equation}

The main application of heat capacity is for studying thermal
stability. Positivity of $C_{Q}$ can guarantee thermal stability
of a system, while its negativity is regarded as an instability.

On the other hand, in order to study the critical behavior of a
thermodynamical system, one is required to obtain an equation of state, $%
P=P(T,V)$. In the context of black hole thermodynamics, one may
find the temperature of a typical black hole in the presence of
the cosmological constant $T=T(m,r_{+},\Lambda ,Q, other\;hairs)$.
We can consider the cosmological constant as a dynamical pressure
and take into account the relation between the event horizon
radius ($r_{+}$) and the volume to find an equation of state,
$P=P(T,V)$. Applying the properties of
a critical point in an isothermal $P-V$ diagram (inflection point), one finds $%
\left( \frac{\partial P}{\partial V}\right) _{T}=\left( \frac{\partial ^{2}P%
}{\partial V^{2}}\right) _{T}=0$. This relation helps us to find
the critical points, and the possible phase transitions. This
method very much depends on the value of temperature. It is worth
mentioning that this method is not practical for black holes with
non-spherical horizon in most gravitational theories. In order to
avoid such a restriction, we should use an alternative method for
obtaining the critical values in the extended phase space.

\subsection{van der Waals liquid-gas system (a brief review)}

The van der Waals system is one of the important models for
describing a real liquid--gas system and its critical behavior.
The equation of state of this model is a modification of the ideal
gas equation, and considers the non-zero size of the molecules and
the attraction between them. The van der Waals equation of state
is given by \cite{Kubiznak}
\begin{equation}
\left( P+\frac{a}{v^{2}}\right) \left( v-b\right) =kT(P,v),  \label{Van}
\end{equation}%
where $P$ and $T$ are the pressure and temperature, respectively.
Also, $v$ is the specific volume $v=\frac{V}{N}$, $b$ is a free
parameter related to the non-zero size of the molecules of a
fluid, and $a$ represents the strength of attraction between the
molecules. Here, $k$ is a constant that can be set to one without
loss of generality. Note that setting $a=b=0$ yields the familiar
ideal gas law. Due to the van der Waals-black hole correspondence,
one can use the analogy between the temperature (and hence
equation of state) of the fluid and the temperature of the black
hole. The existence of critical behavior can be determined by
examining the properties of the inflection point, which satisfy
\begin{equation}
\left( \frac{\partial P}{\partial v}\right) _{T}=\left( \frac{\partial ^{2}P%
}{\partial v^2}\right) _{T}=0.  \label{inf}
\end{equation}

Using the inflection point of the equation of state of the van der
Waals system Eq. (\ref{Van}), it is straightforward to find the
following critical values
\begin{equation}
v_{c}=3b,~~~\&~~~P_{c}=\frac{a}{27b^{2}},~~~\&~~~T_{c}=\frac{8a}{27bk}.
\label{Crit}
\end{equation}

Inserting $T=T_{c}$ in the equation of state, we find two
inseparable phases of liquid-gas with a possible phase transition
between them. For the case $T<T_{c}$, we have a phase transition
between two phases of liquid and gas. However, there exists a
region of specific volume in which no physical system exists and
the phase transition takes place over it. In other words, for this
case, two specific volumes with the same pressure exist, and the
phase transition takes place between them. In order to obtain all
of these critical behaviors and their specific critical values,
one has to take into account all temperatures smaller than
$T_{c}$, which is practically impossible. Our method, introduced
in this paper, provides the possibility of obtaining all of these
critical points analytically. We will demonstrate this possibility
in what follows.

The Gibbs free energy of this system can be expressed in the
following form
\begin{equation}
G=-kT\left( 1+\ln \left[ \frac{v-b}{\Phi }T^{\frac{3}{2}}\right] \right) -%
\frac{a}{v}+Pv,  \label{GV}
\end{equation}
where $\Phi$ is a constant characterizing the gas. The entropy of
the system can be obtained from the differential equation
$dG=-SdT+vdP$, which leads to
\begin{equation}
S=k\left( \frac{5}{2}+\ln \left[ \frac{v-b}{\Phi }T^{\frac{3}{2}}\right]
\right) .  \label{S(v,T)}
\end{equation}

Using the equation of state and inserting the temperature in Eq. (%
\ref{S(v,T)}), one can find the following $S(v,P)$
\begin{equation}
S(v,P)=k\left( \frac{5}{2}+\ln \left[ \frac{\left( v-b\right) ^{\frac{5}{2}%
}\left( P+\frac{a}{v^{2}}\right) ^{\frac{3}{2}}}{k^{\frac{3}{2}}\Phi }\right]
\right) .  \label{S(v,P)}
\end{equation}

In order to use the method introduced in this paper, one needs the
enthalpy of the system. This quantity can be calculated using
different methods, and the following relation is one of them
\begin{equation}
H=G+TS=\frac{3}{2}kT-\frac{a}{v}+Pv,  \label{HV}
\end{equation}
which is known as the enthalpy of the van der Waals system. Using
the equation of state, it is easy to find the following relation
for $H(v,P)$
\begin{equation}
H(v,P)=\left( \frac{5v-3b}{2}\right) P+\frac{a\left( v-3b\right) }{2v^{2}},
\label{H(v,P)}
\end{equation}
where, in principle, one may remove $v$ from Eqs. (\ref{S(v,P)})
and (\ref{H(v,P)}) to obtain $H=H(S,P)$.

\section{Alternative approach to phase transitions \label{NewProp}}

Taking into account the postulates of the usual thermodynamics, it
appears that all complete differentiations can be written as
functions of three thermodynamic variables. It is known that these
three variables are not independent, for instance, an equation of
state can reduce the number of degrees of freedom to two. On the
other hand, the equations for the thermodynamical properties of
the system are combinations of different variables. For example,
in most cases, pressure and temperature, Gibbs free energy and
internal energy, etc., of a thermodynamical system are not
independent of each other. Pressure and temperature are related by
the equation of state, and the Gibbs free energy may be derived by
the Legendre transform of the internal energy. Hence, internal and
Gibbs free energies, as pressure and temperature, are dependent on
each other.

In practice, obtaining all possible critical points of a system,
as well as the ranges in which phase transitions take place, is
not possible with the usual methods. The main reason is that one
has to take the value of a particular thermodynamical quantity
smaller than its critical value to obtain the critical points and
their corresponding range of phase transitions. Mathematically, it
is not possible to solve such a problem analytically using the
usual methods. We would like to introduce here an alternative
method that provides such a possibility, and which results in new
relations between thermodynamical quantities. These relations
provide information about the phase transitions of a system and
their corresponding ranges.

Since divergence points of the heat capacity ($\left(
\frac{\partial S}{\partial T}\right) _{P}=\frac{1}{\left(
\frac{\partial T}{\partial S}\right) _{P}} \rightarrow \infty$)
hint us for the possible phase transition (see Eq. (\ref{CQ})),
the equation of vanishing slope of $T$ versus $S$ helps us to
obtain a new relation for pressure (and look for phase transition
by maximizing the new relation for pressure). We should note that
in order to find the critical behavior of a system, the vanishing
slope of $T$ versus $S$ is necessary but not sufficient in our
approach. \emph{The method is as follows:} instead of considering
the usual equation of state, we use the equation of vanishing
slope of $T$ versus $S$, $\left( \frac{\partial T}{\partial
S}\right) _{P}=0$. This equation is solved with respect to
pressure in the extended phase space. This leads to a new relation
for the pressure which is only volume dependent. This relation
differs from the other relations which are obtained using the
usual equation of state. The existence of a maximum of pressure
from this relation must be examined. The maximum(maxima) of this
relation is(are) the critical point(s) where the phase transition
takes place. In other words, the maximum of this relation is where
the system undergoes a phase transition.

It is evident that by finding a maximum, one is able to extract
the critical pressure and the horizon radius (volume) at the same
time. It is worth
mentioning that to obtain the relation for pressure one can use $%
\left( \frac{\partial ^{2}H}{\partial S^{2}}\right) _{P}$ instead
of $\left( \frac{\partial T}{\partial S}\right) _{P}$, where $H$
is the enthalpy of the system. The maximum is where the system
acquires a phase transition (the maximum indicates the critical
values for a system). On the other hand, thermodynamical concepts
indicate that for pressures smaller than the critical one, two
critical volumes exist between which a phase transition takes
place. For pressures larger than the critical pressure, no phase
transition exists. We see that due to the existence of critical
pressure at the maximum of this relation, such a property is
preserved in our method, and all possible critical points and
their corresponding ranges are obtained. Furthermore, using the
new relation for pressure, and replacing it in the usual equation
of state, one can obtain a new relation for temperature, which is
free of pressure. The same can be done for enthalpy, internal
energy or Gibbs free energy, which leads to new relations which
are only volume dependent. In order to elaborate the efficiency of
the presented method, we give in what follows two typical but
general examples in the context of the usual thermodynamics and
black hole thermodynamics.

\subsection{The case of van der Waals liquid/gas system}

We are now in a position to use our method for the case of  van
der Waals system. Since both the entropy $S(v) $ and enthalpy
$H(v)$ are volume dependent at constant pressure (see Eqs.
(\ref{S(v,P)}) and (\ref{H(v,P)})), we can use the following
relation
\begin{equation}
\left( \frac{\partial ^{2}H}{\partial S^{2}}\right) _{P}=\left( \frac{%
\partial S}{\partial v}\right) _{P}^{-1}\frac{\partial }{\partial v}\left[
\left( \frac{\partial H}{\partial v}\right) _{P}\left( \frac{\partial S}{%
\partial v}\right) _{P}^{-1}\right] _{P},
\end{equation}%
which leads to the following relation for a van der Waals
liquid/gas system
\begin{equation}
\left( \frac{\partial ^{2}H}{\partial S^{2}}\right) _{P}=\frac{2\left(
v-b\right) \left( P+\frac{a}{v^{2}}\right) \left( Pv^{3}-av+2ab\right) }{%
k^{2}\left( 5Pv^{3}-av+6ab\right) }.
\end{equation}

From the above relation, one finds that enthalpy changes during a
phase transition. In other words, the increasing/decreasing
behavior of enthalpy is different before and after a phase
transition, and also before and after a critical point. By solving
$\left( \frac{\partial ^{2}H}{\partial S^{2}}\right) _{P}=0$ with
respect to $P$, one can get the following new relation for
pressure, which differs from the usual equation of state
\begin{equation}
P_{new}=\left\{
\begin{array}{cc}
\frac{v-2b}{v^{3}}a, & \text{allowed} \\
-\frac{a}{v^{2}}, & \text{unallowed, due to the equation of state}%
\end{array}%
\right. .  \label{PNV}
\end{equation}

Using the concept of extremum of this (allowed) relation, that is
the critical point, one finds following critical volume and
pressure
\begin{equation}
v_{c}=3b,~~~~\&~~~~P_{c}=\frac{a}{27b^{2}},
\end{equation}%
which are identical to those obtained previously in Eq.
(\ref{Crit}). By replacing the new pressure in the equation of
state, Gibbs free energy and enthalpy, one can get new relations
for these thermodynamical quantities (as well as for other), which
are pressure independent
\begin{eqnarray*}
T_{new} &=&\frac{2a\left( b-v\right) ^{2}}{kv^{3}}, \\
G_{new} &=&-kT\left( 1+2\ln \left[ \frac{v-b}{\Phi }T^{\frac{3}{2}}\right]
\right) -\frac{2ab}{v}, \\
H_{new} &=&\frac{3}{2}kT-\frac{2ab}{v^{2}}.
\end{eqnarray*}

\begin{center}
\begin{figure}[t]
$%
\begin{array}{c}
\epsfxsize=5cm \epsffile{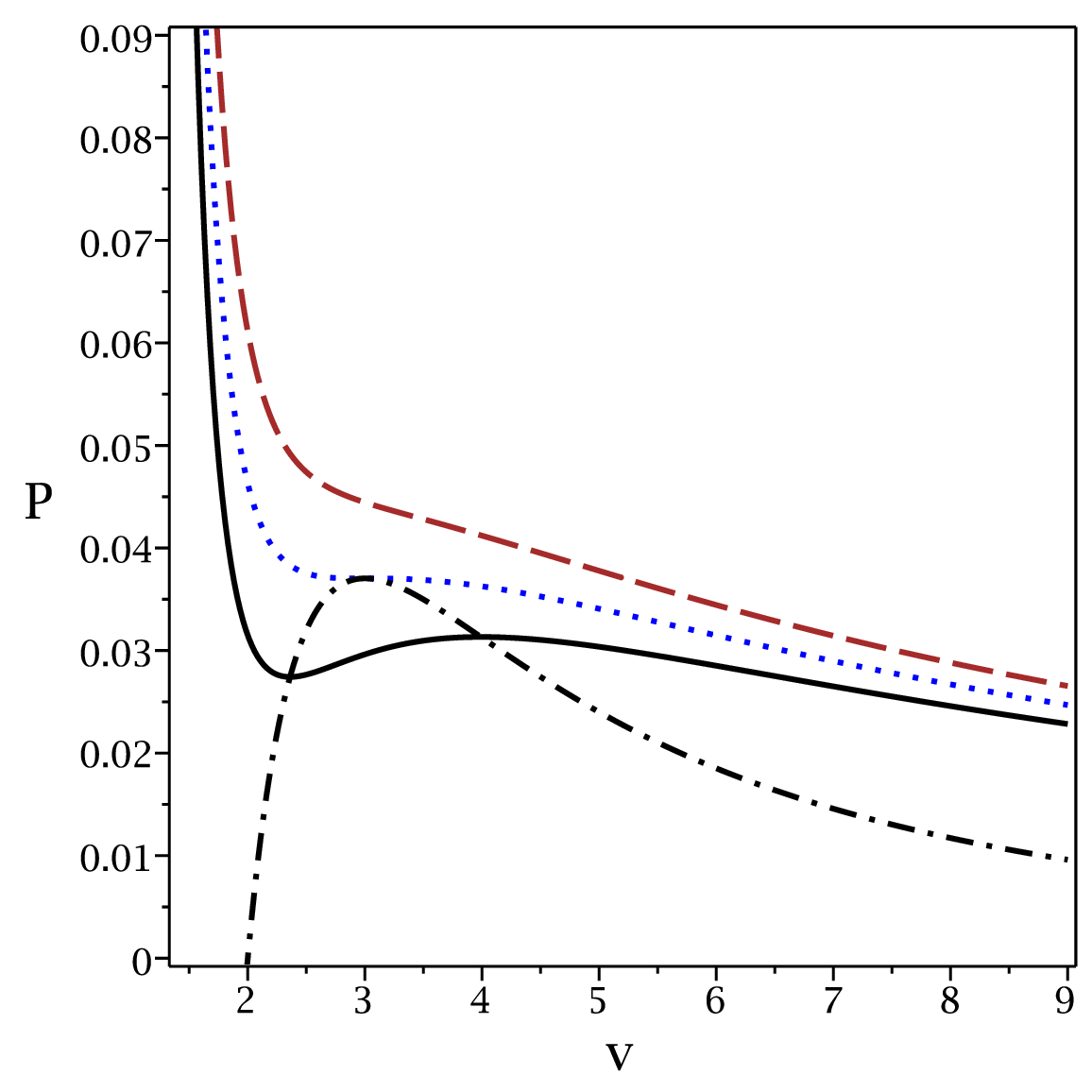} \\
\epsfxsize=5cm \epsffile{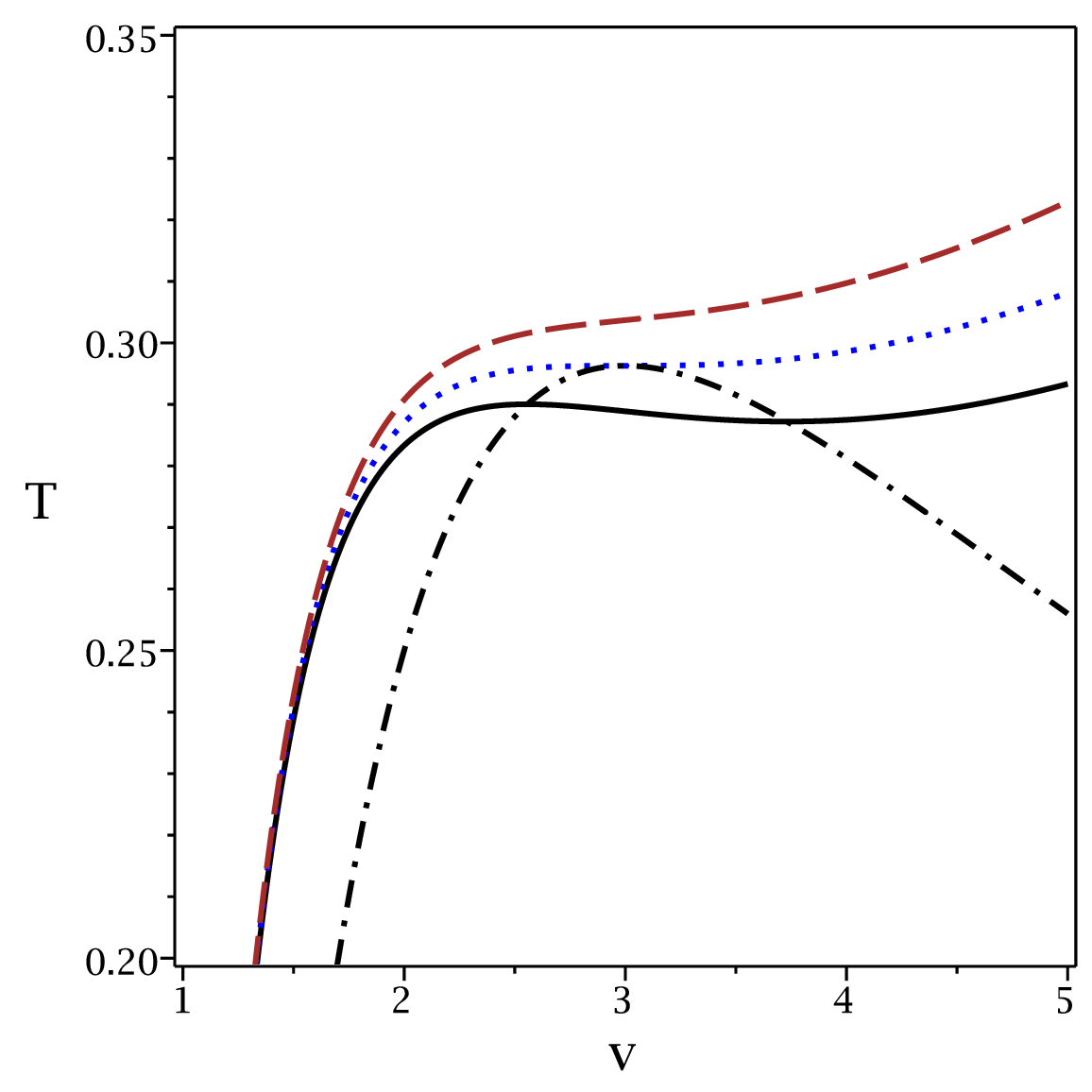}%
\end{array}
$%
\caption{ (color online) Application of the method to the case of
the van der Waals system. \textbf{Top panel:} $P_{new}$
(dash-dotted line) and $P$
versus $v$ for $T=0.9T_{c}$ (continuous line), $T=T_{c}$ (dotted line) and $%
T=1.1T_{c}$ (dashed line). \\[0pt]
\textbf{Bottom panel:} $T_{new}$ (dash-dotted line) and $T$ versus
$v
$ for $P=0.9P_{c}$ (continuous line), $P=P_{c}$ (dotted line) and $%
P=1.1P_{c} $ (dashed line). In both diagrams, we chose $b=1$, $a=4$ and $k=1$%
.}
\label{FigVan}
\end{figure}
\end{center}

These new relations enable us to extract all possible critical
temperatures, Gibbs free energy and enthalpy that a system can
acquire. In order to highlight this aspect of our method, we use
the plot in Fig. \ref{FigVan}.


It is evident that the maxima of the new relations for the
temperature $T_{new}$ and pressure $P_{new}$ are where the system
undergoes a phase transition. Note that in the $P-v$ picture, the
temperature is kept constant, while in the $T-v$ picture, the
pressure is kept constant. In addition, for pressures
(temperatures) smaller than the critical pressure (temperature),
the new relation gives a single pressure (temperature) with two
related volumes. The phase transition takes place between these
two volumes at a specific pressure. As one can see, all possible
critical points and the corresponding ranges of phase transitions
are included in this method. This is one of the important features
of our method which was not
possible with previous methods. It is interesting to note that the minima of $%
T $ ($P$) coincide with the maxima of the $T_{new}$ ($P_{new}$).
We continue with another example in the context of black holes.

\section{Phase transition in higher dimensional Reissner-Nordstr\"{o}m AdS
Black hole \label{RN}}

The main motivation to study asymptotically AdS black holes stems
from the hypothesis of AdS/CFT correspondence. Using the thermal
field theoretic approaches, it has been deduced that AdS black
holes undergo certain phase transitions. The first sign of such
phase transitions was observed by Page and Hawking for the
Schwarzschild AdS black hole \cite{hawking}. With the addition of
parameters such as electric charge and spin, the phase transition
process is more elaborate and enhanced. It is quite interesting to
note that the pressure-volume picture of the ideal gas for
constant temperature is also mimicked by the AdS black holes (see
Fig.1 top panel). This analogy between a gravitational system (an
AdS black hole) and a non-gravitational thermal system (such as an
ideal gas or a van der Waals fluid) is established by identifying
a correspondence between their parameters i.e. mass with enthalpy,
temperature with surface gravity, entropy with area, and
cosmological constant with pressure. Thus, in the first law of
thermodynamics, the cosmological constant appears as pressure,
which is conjugate to the volume of a black hole \cite{mann}.
Besides, using the reverse isoperimetric inequality, it has been
deduced that entropy inside the horizon of a given volume is
maximized for the Schwarzschild AdS black hole \cite{miriam}.

In the black hole systems, it was shown that one can take the
negative cosmological constant as thermodynamical pressure
\cite{dolan} with the following relation
\begin{equation}
P=-\frac{\Lambda }{8\pi }.  \label{P}
\end{equation}

On a complementary note, we should mention that for specific black
holes in modified general relativity, such as dilatonic gravity
and gravity's rainbow, one has to use a modified proportionality
relation instead of Eq. (\ref{P}) \cite{HendiFEP,HendiPEFM}.
Although in this paper we consider a well-known
Reissner-Nordstr\"{o}m AdS black hole, our technique is consistent
with the other black holes in modified theories of gravity.
Replacing the cosmological constant with thermodynamical pressure
(working in the extended phase space thermodynamics) leads to the
following important results:

I) The resulting temperature for the black hole is the equation of state.

II) The total mass of the black hole is no longer the internal
energy. In fact, it is replaced by the enthalpy in such
considerations. This results in the following relation for the
Gibbs free energy
\begin{equation}
G=M-TS.  \label{Gibbs1}
\end{equation}

We would like to give now an example of the validity of our
approach and its consistency with previous methods in the context
of black hole systems. For this purpose, we study the critical
behavior of the $d-$dimensional Reissner-Nordstr\"{o}m AdS black
hole. Previously, the results for this specific black hole were
derived using the usual method in Ref. \cite{Kubiznak}. The metric
of this black hole in spherically symmetric spacetime is given as
\begin{equation}
ds^{2}=-\psi(r) dt^{2}+\frac{dr^{2}}{\psi(r)}+ r^2 d \Omega _{d_{2}}^2,
\label{metric}
\end{equation}
where we use the notation $d_{i}=d-i$. In Eq. (\ref{metric}), $d
\Omega _{d_{2}}^2$ denotes the metric of $(d_{2})-$dimensional
unit sphere, and
\begin{equation}
\psi(r) =1-\frac{m}{r^{d_{3}}}-\frac{2\Lambda }{d_{1}d_{2}}r^{2}+\frac{%
2d_{3}q^{2}}{d_{2}r^{2d_{3}}}.  \label{f(r)}
\end{equation}

The temperature, entropy and total finite mass of this black hole
are calculated using the surface gravity, area law and ADM
approach, respectively, which lead to
\begin{eqnarray}
T &=&\frac{d_{3}}{4\pi r_{+}}-\frac{\Lambda r_{+} }{2\pi d_{2}}-\frac{%
d_{3}^{2}q^{2}}{2\pi d_{2}r_{+}^{2d_{5/2}}},  \label{TotalTT} \\
S &=&\frac{r_{+}^{d_{2}}}{4},  \label{TotalS} \\
M &=&\frac{d_{2}}{16\pi }m,  \label{TotalM}
\end{eqnarray}
where $r_{+}$ is the outer horizon of the black hole. By
evaluating the metric function on the outer horizon ($\psi \left(
r=r_{+}\right) =0$), we obtain
\begin{equation}
M=\frac{d_{2}}{16\pi }\left( r_{+}^{d_{3}}-\frac{2r_{+}^{d_{1}}}{d_{1}d_{2}}%
\Lambda +\frac{2d_{3}q^{2}}{d_{2}r_{+}^{d_{3}}}\right),  \label{Mass}
\end{equation}
with the following relation for the Gibbs free energy
\begin{equation}
G=\frac{r_{+}^{d_{3}}}{16\pi }+\frac{d_{3}\left( 2d-5\right) q^{2}}{8\pi
d_{2}r_{+}^{d_{3}}}-\frac{r_{+}^{d_{1}}}{d_{1}d_{2}}P.  \label{Gibbs}
\end{equation}


Another interesting method for calculating the thermodynamic
potentials (such as the Gibbs free energy) in a gravitational
system is based on the Euclidean on-shell action. Since bulk
action of the theory diverges, we use the counter-term action (as
a boundary action) to remove the divergency. In addition, we
should add the Gibbons-Hawking and electromagnetic boundary terms
to the bulk action to get a well-defined action. The well-behaved
finite action can be written as (see \cite{DEhShakVah})
\begin{equation}
I=I_{b}+I_{ct}-\frac{1}{8\pi }\int_{\partial M}d^{n}x~\sqrt{\gamma }~K-\frac{%
1}{4\pi }\int_{\partial M}d^{n}x~\sqrt{\gamma }n_{\mu }F^{\mu \nu }A_{\nu },
\label{FullAction}
\end{equation}
where $I_{b}$ and $I_{ct}$ are, respectively, the bulk and
counter-term actions of the Einstein-Maxwell gravity. Also,
$\gamma _{ij}$ and $K$ are, respectively, the induced metric and
the extrinsic curvature of the boundary. Using Eq.
(\ref{FullAction}), it is straightforward to calculate the total
on-shell action with respect to the volume of the unit
$d_{2}-$sphere
\begin{equation}
I=\frac{\beta }{16\pi }\left( r_{+}^{d_{3}}+{\frac{2\Lambda r_{+}^{d_{1}}}{%
d_{1}d_{2}}}+\frac{2d_{3}(2d-5)q^{2}}{d_{2}r_{+}^{d_{3}}}\right) ,
\label{Onshell}
\end{equation}
where $\beta$ is the inverse of the Hawking temperature. Using the fact that $%
G=I/\beta$, with Eq. (\ref{P}) we find
\begin{equation}
G=\frac{1}{16\pi }\left( {r}_{+}^{d_{3}}-\frac{16\pi Pr_{+}^{d_{1}}}{%
d_{1}d_{2}}+\frac{2d_{3}(2d-5)q^{2}}{d_{2}r_{+}^{d_{3}}}\right),
\label{Gibbs2}
\end{equation}
which is the same as Eq. (\ref{Gibbs}), as expected.

We are now in a position to calculate the critical values with the
usual method. First, we calculate the volume conjugate to the
pressure as
\begin{equation}
V=\left( \frac{\partial H}{\partial P}\right) _{S,Q}=\left( \frac{\partial M%
}{\partial P}\right) _{S,Q}=\frac{r_{+}^{d_{1}}}{d_{1}}.  \label{V}
\end{equation}

Since the volume depends on the horizon radius, one can use the
horizon radius for investigating the thermodynamic behavior of the
black hole, proportional (linearly) to its specific volume
\cite{Kubiznak}. Using Eqs. (\ref{P}) and (\ref{TotalTT}), one
obtains the equation of state as
\begin{equation}
P=\frac{d_{2}\left( 4\pi Tr_{+}^{3}-d_{3}r_{+}^{2}\right) }{16\pi r_{+}^{4}}+%
\frac{2d_{3}^{2}q^{2}r_{+}^{-2d_{4}}}{16\pi r_{+}^{4}}.  \label{PP}
\end{equation}

We now employ the proprieties of the inflection point, $\left( \frac{\partial P%
}{\partial r_{+}}\right) _{T}=\left( \frac{\partial ^{2}P}{\partial r_{+}^{2}%
}\right) _{T}=0$, to obtain the critical horizon radius (volume)
$r_{c}$, which leads to
\begin{equation}
r_{c}^{2}-4d_{3}d_{5/2}q^{2}r_{c}^{-2d_{4}}=0
\end{equation}%
with the following solution \cite{Kubiznak}
\begin{equation}
r_{c}=\left( 4 {q}^{2}d_{3}d_{5/2}\right) ^{\frac{1}{2d_{3}}}.  \label{rcc}
\end{equation}

It is a matter of calculation to obtain the critical temperature and
pressure as
\begin{eqnarray}
T_{c} &=&\frac{d_{3}}{2\pi \left( 4{q}^{2}d_{3}d_{5/2}\right) ^{\frac{1}{%
2d_{3}}}}-\frac{\left( 4{q}^{2}d_{3}d_{5/2}\right) ^{\frac{d_{5/2}}{d_{3}}%
}d_{3}^{2}{q}^{2}}{\pi },  \label{Tcc} \\
&&  \notag \\
P_{c} &=&\frac{d_{2}d_{3}}{16\pi \left( 4{q}^{2}d_{3}d_{5/2}\right) ^{\frac{1%
}{d_{3}}}}-\frac{d_{5/2}d_{3}^{2}{q}^{2}}{4\pi \left( 4{q}%
^{2}d_{3}d_{5/2}\right) ^{\frac{d_{2}}{d_{3}}}}.  \label{Pcc}
\end{eqnarray}

Let us now get the critical values using our approach. Using Eqs.
(\ref{TotalS}) and (\ref{Mass}), and replacing the cosmological
constant with the pressure in Eq. (\ref{P}), one can obtain (since
at constant pressure both $S=S(r_{+})$ and $M=M(r_{+})$ are
independent of the temperature, it is not needed to use the
equation of state for removing $T$, which was used before in the
van der Waals liquid/gas system)
\begin{equation}
\left( \frac{\partial ^{2}M}{\partial S^{2}}\right) _{q,P}=\frac{16P}{%
d_{2}^{2}r_{+}^{d_{3}}}-\frac{d_{3}}{\pi d_{2}r_{+}^{d_{1}}}+\frac{4 d_{5/2}
d_{3}^{2} q^{2}}{\pi d_{2}^{2} r_{+}^{3d_{7/3}} }.  \label{CQQ}
\end{equation}

Solving $\left( \frac{\partial ^{2}M}{\partial S^{2}}\right)
_{q,P}=0$ with respect to $P$, we obtain a new relation for the
pressure
\begin{equation}
P_{new}={\frac{d_{2}d_{3}}{16\pi {r}_{+}^{2}}}-\,{\frac{d_{5/2}d_{3}^{2}
q^{2}}{4\pi {r}_{+}^{2d_{2}}}}.  \label{Pnew}
\end{equation}

Replacing the pressure in the relations for
temperature Eq. (\ref{TotalTT}), mass Eq. (\ref{Mass}) and Gibbs free energy Eq. (\ref%
{Gibbs}) with the new pressure relation (\ref{Pnew}), one can
obtain new relations for these thermodynamical quantities in
following form
\begin{eqnarray}
T_{new} &=&{\frac{d_{3}}{2\pi {r}_{+}}}-{\frac{d_{3}^{2}{r}_{+}^{5-2d}q^{2}}{%
\pi }},  \label{Tnew} \\
&&  \notag \\
M_{new} &=&{\frac{d_{2}^{2}{r}_{+}^{d_{3}}}{8\pi d_{1}}}-{\frac{%
d_{2}d_{3}d_{4}q^{2}}{4\pi d_{1}{r}_{+}^{d_{3}}}},  \label{Mnew} \\
&&  \notag \\
G_{new} &=&{\frac{{r}_{+}^{d_{3}}}{8\pi d_{1}}}+{\frac{d_{5/2}d_{3} q^{2}}{2
\pi d_{1}{r}_{+}^{d_{3}}}}.  \label{Gnew}
\end{eqnarray}


\begin{center}
\begin{figure}[t]
$%
\begin{array}{c}
\epsfxsize=5cm \epsffile{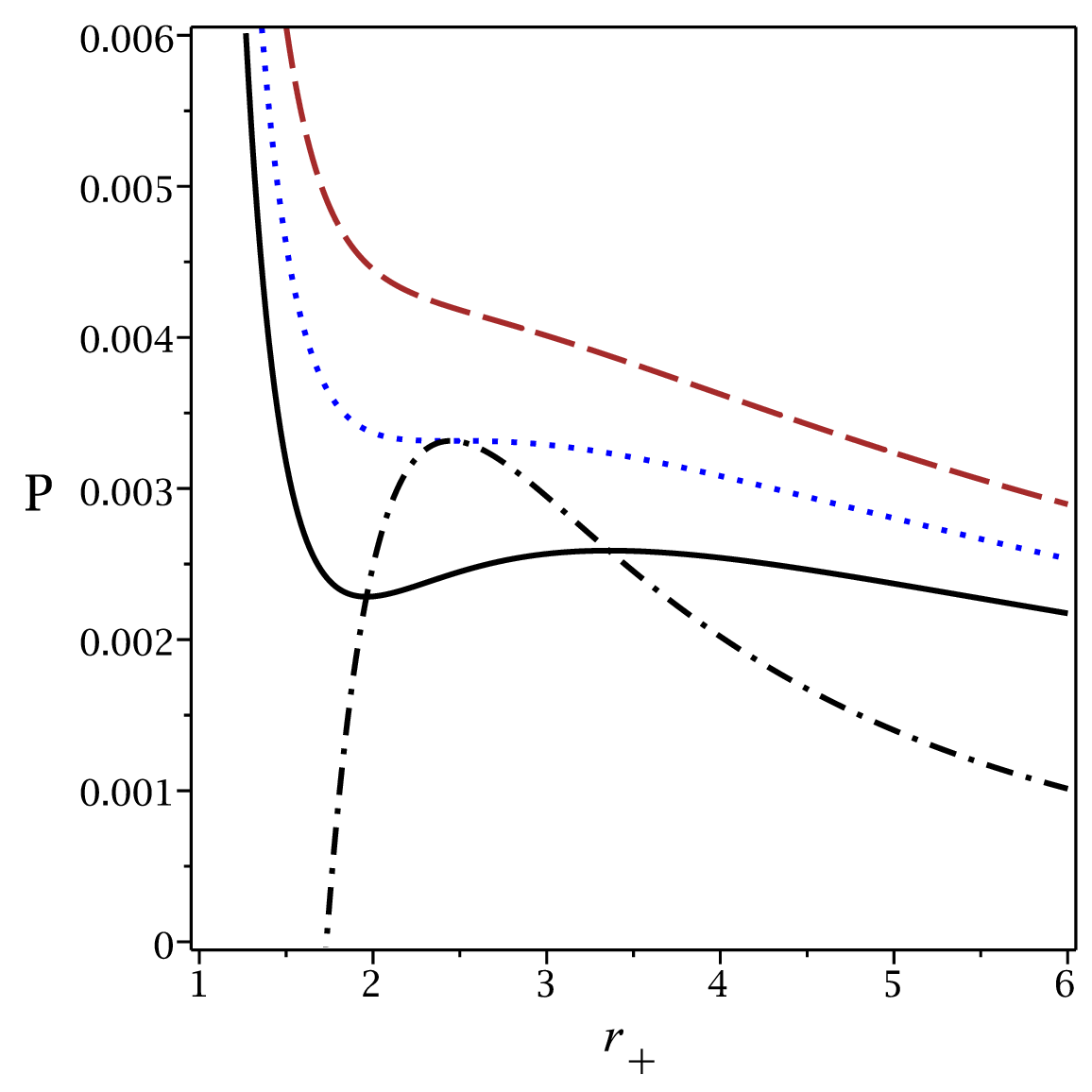} \\
\epsfxsize=5cm \epsffile{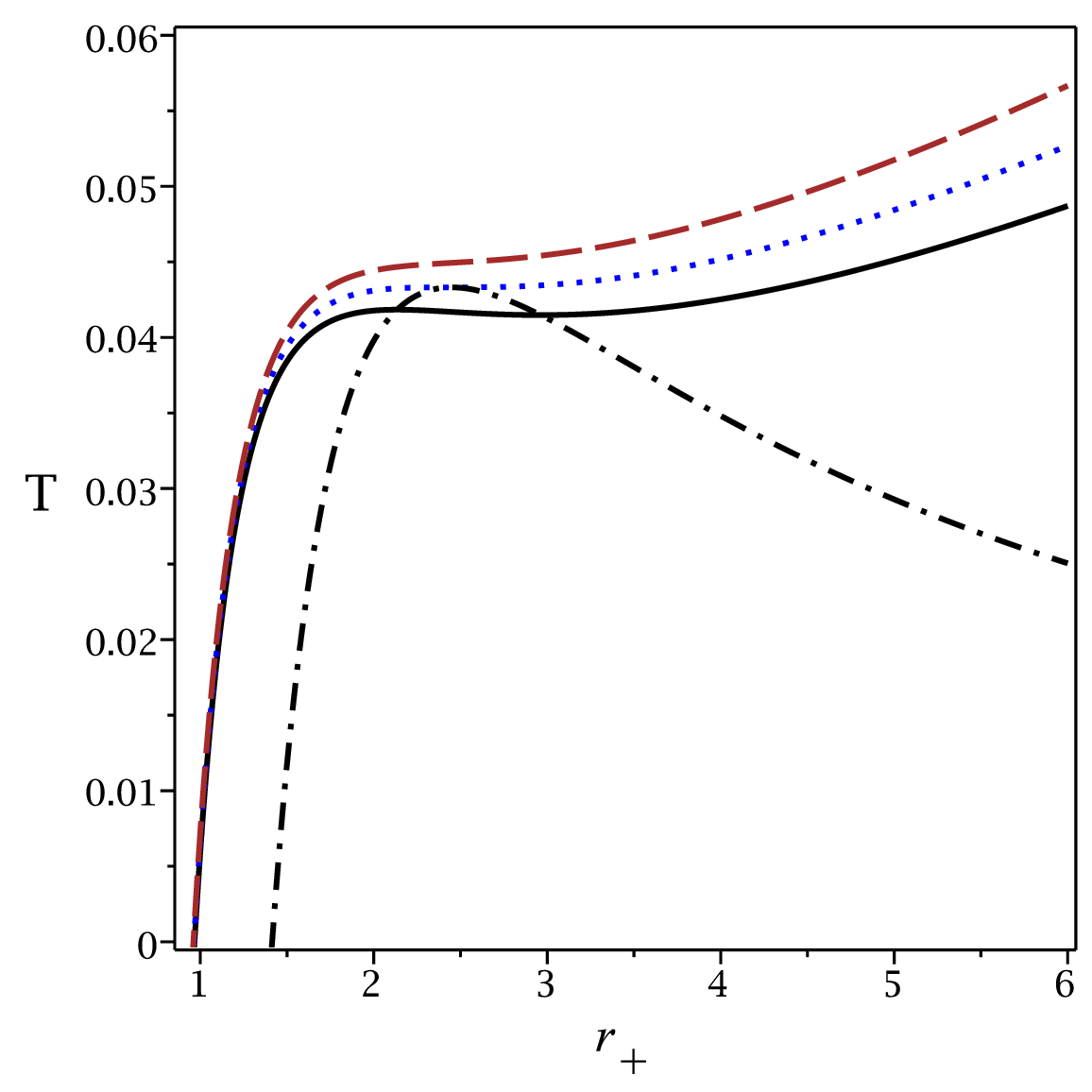} \\
\epsfxsize=5cm \epsffile{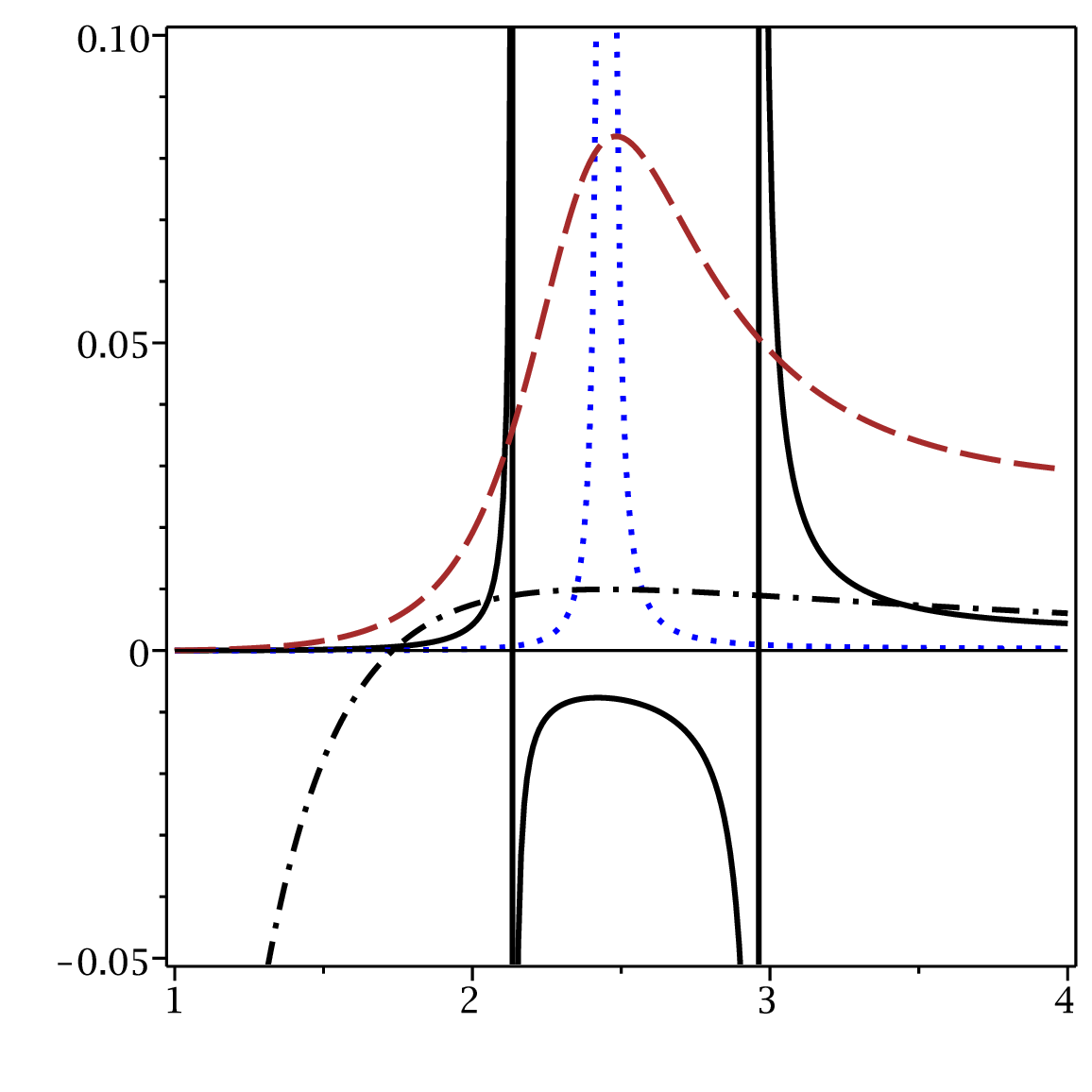}%
\end{array}
$%
\caption{(color online) \textbf{Top panel:} $P_{new}$ (dash-dotted
line) and $P$ versus $r_{+}$, for sub-critical case $T=0.9T_{c}$
(continuous line), critical case $T=T_{c}$ (dotted line) and
super-critical case $T=1.1T_{c}$
(dashed line). \\[0pt]
\textbf{Middle panel:} $T_{new}$ (dash-dotted line) and $T$ versus $%
r_{+}$, for sub-critical $P=0.9P_{c}$ (continuous line), critical $P=P_{c}$
(dotted line) and super-critical $P=1.1P_{c}$ (dashed line). \\[0pt]
\textbf{Bottom panel:} $P_{new}$ (dash-dotted line) and $C_{Q}$
versus $r_{+}$, for sub-critical $P=0.9P_{c}$ (continuous line), critical $%
P=P_{c}$ (dotted line) and super-critical $P=1.1P_{c}$ (dashed
line). In all three panels, $q=1$ and $d=4$.} \label{Fig1-2}
\end{figure}
\end{center}


It is evident that the relation for pressure Eq. (\ref{Pnew}) is
different from the usual equation of state Eq. (\ref{PP}). In
order to obtain the maximum of this relation, we use the
mathematical nature of the extremum
\begin{equation}
\left. \left( \frac{dP_{new}}{dr_{+}}\right) \right\vert _{r_{+}=r_{_{NC}}}=%
\frac{d_{2}d_{3}^{2}d_{5/2}{q}^{2}}{2 \pi r_{_{NC}}^{2d_{3/2}}}-\frac{%
d_{2}d_{3}}{8\pi r_{_{NC}}^{3}}=0
\end{equation}
with the following solution for the new critical (NC) horizon radius, $%
r_{_{NC}}$
\begin{equation}
r_{_{NC}}=\left( 4 d_{3}d_{5/2} {q}^{2} \right) ^{\frac{1}{2d_{3}}},
\label{rcc1}
\end{equation}
which is exactly the same as that obtained in Eq. (\ref{rcc}).
Replacing this horizon radius in Eqs. (\ref{Tnew}) and
(\ref{Pnew}), it is a matter of calculation to find
\begin{eqnarray}
T_{_{NC}} &=&\frac{d_{3}}{2\pi \left( 4{q}^{2}d_{3}d_{5/2}\right) ^{\frac{1}{
2d_{3}}}}-\frac{\left( 4{q}^{2}d_{3}d_{5/2}\right) ^{\frac{d_{5/2}}{d_{3}}%
}d_{3}^{2}{q}^{2}}{\pi }, \\
&&  \notag \\
P_{_{NC}} &=&\frac{d_{2}d_{3}}{16\pi \left( 4 {q}^{2}d_{3}d_{5/2}\right) ^{%
\frac{1}{d_{3}}}}-\frac{d_{5/2}d_{3}^{2}{q}^{2}}{4\pi \left( 4 q
^{2}d_{3}d_{5/2}\right) ^{\frac{d_{2}}{d_{3}}}},
\end{eqnarray}
which are exactly the same as the previously calculated critical
temperature Eq. (\ref{Tcc}) and pressure Eq. (\ref{Pcc}). These
results show that the critical values calculated in our approach
are consistent with those calculated with the usual method in
extended phase space.

Using the critical radius Eq. (\ref{rcc1}) with the new relations
for the mass Eq. (\ref{Mnew}) and Gibbs free energy Eq.
(\ref{Gnew}), we obtain the critical mass (enthalpy) and critical
Gibbs free energy as well
\begin{eqnarray*}
M_{_{NC}}&=&\frac{d_{2}d_{3}\left( d^{2}-5d+7\right) q}{4\pi d_{1}\sqrt{%
d_{3}d_{5/2}}}, \\
G_{_{NC}}&=&\frac{\sqrt{d_{3}d_{5/2}}q}{2\pi d_{1}}.
\end{eqnarray*}

In order to elaborate the results of our approach in more detail,
we plot in Fig. \ref{Fig1-2} the obtained relations for
temperature and pressure. It is clear that for pressures larger
than the critical one (the temperature is larger than its critical
value in the van der Waals-like diagram), no phase transition is
observed for the van der Waals-like diagram (dashed line of Fig.
\ref{Fig1-2}, top and middle panels). If no phase transition
occurs in a black hole, this implies that the black hole remains
physically intact, i.e. its mass and the other physical parameters
remain the same. The black hole remains stable and does not
radiate thermally. This may correspond to a state of thermal
equilibrium. If the equilibrium gets unstable, the heat capacity
of the black hole become negative, causing the black hole to
radiate. In this case, a phase transition does take place.
Similarly for same conditions, no critical pressure is observed in
our approach. On the contrary, for pressures smaller than the
critical value (the temperature is smaller than the critical
temperature), two critical horizons are observed for any pressure
(temperature) which is in agreement with the results of our
approach (continuous line in Fig. \ref{Fig1-2}, top and middle
panels). Finally, we observe that the critical pressure (critical
temperature) and the critical horizon radius calculated by the
usual method, coincide with the maximum of the new relation for
pressure and its related horizon radius. This also indicates that
the results of our method, are completely in agreement with the
previous method.

Finally, we plot the heat capacity (bottom panel in Fig.
\ref{Fig1-2}) to show the consistency of the new pressure. It is
evident that for $P<P_{c}$, two points of discontinuity exist for
the heat capacity, which are coincident with the phase transition
points that are observed in the other methods. If the pressure is
equal to the critical pressure, only one discontinuity is observed
in the heat capacity, as in the other methods. For $P>P_{c}$, no
discontinuity is observed for the heat capacity. This behavior
indicates that all methods give consistent results.

In the top panel of Fig. \ref{Fig1-2}, one can see the so-called
saturation curve, dash-dotted line. Taking into account the $P-V$
isothermal diagram with $T<T_{c}$ (continuous line), we can
decrease the horizon radius to find two points of intersection
with the saturation curve ($r_{+1}$ and $r_{+2}$ with
$r_{+1}<r_{+2}$). The black hole system is unstable for $%
r_{+1}<r_{+}<r_{+2}$. In other words, there is a phase transition
between a small and a large black hole (between $r_{+1}$ and
$r_{+2}$). This phase transition may occur with a sudden burst of
thermal Hawking radiations, i.e. the size of the black hole
suddenly shrinks from $r_{+2} $ to $r_{+1}$ without changing the
black hole temperature. Black hole solutions are not physical
between these two points. This can be explained
by the fact that the heat capacity is negative (see bottom panel of Fig. \ref%
{Fig1-2}, for more detail), and also by the fact that the speed of
sound is larger than the speed of light \cite{Riazi}. Note that
similar discontinuities in specific heat capacity also occur in
the Born-Infeld black holes \cite{r2}. It is worth noting that for
$T=T_{c}$, the intersection points meet and are equal to the
critical horizon radius $r_{+1}=r_{+2}=r_{c}$. The same statement
could be made for the temperature in the middle panel of Fig.
\ref{Fig1-2}.

Before finishing the paper, it is worth pointing out the
significance of our approach.

\emph{First} of all, our method provides the possibility of
obtaining different thermodynamical quantities independent of each
other. In other words, as one can see from Eqs.
(\ref{Tnew})-(\ref{Gnew}), they only depend on such properties as
dimension, electric charge and horizon radius. If we generalize
the action to other gravitational theories or include other matter
fields, the resultant new temperature, pressure, mass and Gibbs
free energy obtained using our method, have the same properties
(they are only a function of black hole properties).

\emph{Second}, the new relations include only critical points that
a black hole could acquire in different conditions. In the usual
methods, in order to get all points between which phase
transitions take place, one has to consider pressures which are
equal or smaller than the critical pressure. Technically, such a
task is impossible if it is to be done for all pressures. Using
our method, one can find all possible phase transitions, horizon
radii and corresponding pressures that a system could acquire. The
same could be said for the new relations for temperature, mass and
Gibbs free energy. In other words, by using our approach, one can
obtain all phase transition points and corresponding critical
temperature, pressure, mass and Gibbs free energy that system can
acquire in analytical form.

\emph{Third}, using our method, one can get the range of horizon
radii that depends on the critical values, in which the black hole
solutions do not exist. For clarification, one should take a look
at the diagram for the new pressure in the top panel of Fig.
\ref{Fig1-2} (dashed-dotted line). Clearly, the phase transition
takes place between two points with the same pressure. The
prohibited range of horizon radii for the black hole is between
these points. Taking a closer look, one can see that by using our
approach, one can obtain the maximum range of the horizon radii in
which the black hole solutions do not exist. Such a maximum could
not be obtained  easily with the usual methods. In addition, by
using our approach, one can get the rate of increase of the
prohibited range of horizon radii by studying the behavior of its
diagram. Such a procedure may encounter significant problems for
the usual method. Finally, we should point out that these three
features are also valid for the usual thermodynamical systems.

\section{Closing remarks}

Motivated by the interest in van der Waals-like behavior and
recent progress in the thermodynamic phase transitions of black
holes, we introduced an alternative approach for studying the
phase transition points in both the usual thermodynamical systems
and black holes.

Although the usual method of studying a phase transition
originates in the temperature as the equation of state, our
approach is based on the slope of temperature versus entropy, and
is a powerful method to address the critical behavior of a
thermodynamical system.

The results of our approach are in agreement with the other
methods. However, it also provides further information regarding
the critical behavior of thermodynamical systems, which could not
be derived with the other methods. As the highlights of our
method, one can state: I) Obtaining new relations for different
thermodynamical quantities which are independent of each other.
II) Mapping all possible critical points and regions in which
phase transitions take place.

Since it is known that the four-dimensional Reissner-Nordstr\"{o}m
black hole has the reentrant phase transition, it would be
interesting to examine if our method confirms (or not) the
existence of the reentrant phase transition. The method introduced
here is applicable to both the usual thermodynamical systems and
black holes. This shows that one can also employ the general
structure of this method in the context of other physical systems
such superconductors, condense matter systems, gauge/gravity
duality, or even in the context of quantum systems. It is also
interesting to build a geometrical theory based on the Legendre
invariance, like the known theory  of geometrothermodynamics, or
on other types of symmetries. As a future task, it would be
interesting to extend the present study of phase transitions and
critical phenomenon via AdS/CFT \cite{ads}. Furthermore, how do
phase transitions alter the geometry and topology of black holes,
is a question for a separate investigation. We also plan to
investigate the relationship between the cosmic censorship
hypothesis and the various phase transition of black holes, and to
work on phase transitions and critical behavior of black rings,
black saturns and black membranes.

\section*{Acknowledgements}

We thank an anonymous referee for useful comments. We also thank
the editor of Chinese Physics C for improving the text and
presentation. We wish to thank the Shiraz University Research
Council. This work has been supported financially by Research
Institute for Astronomy and Astrophysics of Maragha (RIAAM), Iran.


\begin{thebibliography}{99}
\bibitem{bio} D. Y. Lando and V. B. Teif, J. Biomol. Struct. Dynam. \textbf{17}, 903 (2000).

\bibitem{nice} G. Longo, and M. Montevil, Progress Biophys. Molec. Biology,
\textbf{106}, 340 (2011).

\bibitem{El-Par} H. Kleinert, Phys. Rev. D \textbf{60}, 085001 (1999).

\bibitem{Con-MatI} A. L. Greer, Science, \textbf{267}, 1947 (1995).

\bibitem{Con-MatII} K. Kumar, A. K. Pramanik, A. Banerjee, P. Chaddah, S. B.
Roy, S. Park, C. L. Zhang and S. W. Cheong, Phys. Rev. B \textbf{73}, 184435
(2006).

\bibitem{Con-MatIII} G. Tarjus, Nature \textbf{448}, 758 (2007).

\bibitem{Thermo} H. B. Callen, "\emph{Thermodynamics and an Introduction to
Thermostatistics}" (John Wiley \& Sons, Inc., New York, 1985).

\bibitem{Cosmol} D. Layzer, "\emph{Cosmogenesis, The Development of Order in
the Universe}", (Oxford Univ. Press, 1991).

\bibitem{BHs} D. C. Zou, Y. Liu and R. Yue, Eur. Phys. J. C \textbf{77}, 365
(2017).

\bibitem{r1} G. J. Stephens and B. L. Hu, Int. J. Theor. Phys. \textbf{40},
2183 (2001).

\bibitem{r2} D. Roychowdhury, [arXiv:1403.4356].

\bibitem{r3} V. Niarchos, Mod. Phys. Lett. A \textbf{23}, 2625 (2008).

\bibitem{r4} T. Harmark, N. A. Obers, [arXiv: hep-th/0503020].

\bibitem{PoincareI} H. Poincare, Acta. Math. \textbf{7}, 259 (1885).

\bibitem{PoincareII} R. D. Sorkin, Astrophys. J. \textbf{257}, 847 (1982).

\bibitem{PoincareIII} O. Kaburaki, I. Okamoto and J. Katz, Phys. Rev. D
\textbf{47}, 2234 (1993).

\bibitem{PoincareIV} M. Azreg-A\"{\i}nou and M. E. Rodrigues, JHEP \textbf{09}, 146 (2013).

\bibitem{Kubiznak} D. Kubiznak and R. B. Mann, JHEP \textbf{07}, 033 (2012).

\bibitem{hawking} S. W. Hawking and D. N. Page, Commun. Math. Phys. \textbf{87},
577 (1983).

\bibitem{mann} A. Rajagopal, D. Kubiznak and R. B. Mann, Phys. Lett. B
\textbf{737}, 277 (2014).

\bibitem{miriam} M. Cvetic, G. W. Gibbons, D. Kubiznak and C. N. Pope, Phys.
Rev. D \textbf{84}, 024037 (2011).

\bibitem{dolan} B. P. Dolan, Class. Quantum Gravit. \textbf{28}, 125020
(2011).

\bibitem{HendiFEP} S. H. Hendi, M. Faizal, B. Eslam Panah and S. Panahiyan,
Eur. Phys. J. C \textbf{76}, 296 (2016).

\bibitem{HendiPEFM} S. H. Hendi, S. Panahiyan, B. Eslam Panah, M. Faizal and
M. Momennia, Phys. Rev. D \textbf{94}, 024028 (2016).

\bibitem{Riazi} S. H. Hendi, N. Riazi and S. Panahiyan, Ann. Phys. (Berlin) \textbf{530}, 1700211
(2018).

\bibitem{DEhShakVah} M. H. Dehghani, Ch. Shakuri and M. H. Vahidinia, Phys.
Rev. D \textbf{87}, 084013 (2013).

\bibitem{ads} M. Natsuume, Prog. Theor. Phys. Suppl. \textbf{186}, 491 (2010).

\end{thebibliography}
\end{document}